\documentclass[12pt, preprint]{aastex}

\shorttitle{ Witnessing Group and BCG Formation }

\shortauthors{Jeltema et al.}

\begin{document}

\title{ RXJ1648.7+6109: Witnessing the Formation of a Massive Group/Poor Cluster and its Brightest Galaxy }

\author{Tesla E. Jeltema\altaffilmark{1}, John S. Mulchaey\altaffilmark{2}, and Lori M. Lubin\altaffilmark{3}}

\altaffiltext{1}{Morrison Fellow, UCO/Lick Observatories, 1156 High St., Santa Cruz, CA 95064; tesla@ucolick.org}
\altaffiltext{2}{The Observatories of the Carnegie Institution of Washington, 813 Santa Barbara St., Pasadena, CA 91101}
\altaffiltext{3}{Department of Physics, University of California at Davis, One Shields Ave., Davis, CA 95616}

\begin{abstract}
Using deep Chandra and optical spectroscopic observations, we investigate an intriguing, young massive group, RXJ1648.7+6109, at $z=0.376$, and we combine these observations with previous measurements to fit the scaling relations of intermediate-redshift groups and poor clusters.  RXJ1648 appears to be in an early stage of formation; while it follows X-ray scaling relations, its X-ray emission is highly elongated and it lacks a central, dominant BCG.  Instead, RXJ1648 contains a central string of seven bright galaxies, which have a smaller velocity dispersion, are on average brighter, and have less star formation (lower EW([OII]) and EW($H_{\delta}$)) than other group galaxies. The 4-5 brightest galaxies in this string should sink to the center and merge through dynamical friction by $z=0$, forming a BCG consistent with a system of RXJ1648's mass even if 5-50\% of the light is lost to an intracluster light component (ICL).  The $L_X-T_X$ relation for intermediate-redshift groups/poor clusters is very similar to the low-redshift cluster relation and consistent with the low-redshift group relation.  In contrast, the $L_X-\sigma_v$ and $\sigma_v-T_X$ relations reveal that intermediate-redshift groups/poor clusters have significantly lower velocity dispersions for their X-ray properties compared to low-redshift systems, however the intermediate-redshift relations are currently limited to a small range in luminosity.
\end{abstract}


\section{ INTRODUCTION }

Groups and poor clusters of galaxies are the building blocks of larger scale structures, and they are particularly important environments for diagnosing the origin of the non-gravitational heating of the intracluster medium (ICM)(Balogh et al.~2006) and for galaxy evolution (Zabludoff \& Mulchaey 1998).  For example, at low redshift non-gravitational heating appears to be proportionally more important in the group regime, leading to a steepening of the $L_X-T$ relation for groups versus clusters and excess group entropy above self-similar expectations (e.g.~Helsdon \& Ponman 2000; Ponman et al.~2003).  The evolution in group X-ray scaling relations with redshift can constrain models of non-gravitational heating (Balogh et al.~2006).  In addition, most galaxies in the local universe lie in groups (e.g.~Turner \& Gott 1976), making them important environments in the study of galaxy evolution.  

We have only recently begun to study the evolution of this important environment with redshift (e.g.~Mulchaey et al. 2006; Jeltema et al. 2006,2007; Willis et al. 2005; Pacaud et al. 2007; Wilman et al. 2005a,b; Gerke et al. 2007) and very few groups/poor clusters at even moderate redshifts ($z>0.2$) have both significant X-ray and optical data (Mulchaey et al. 2006; Jeltema et al. 2006,2007; Willis et al. 2005; Gastaldello et al. 2007). Current X-ray observations at intermediate-redshifts ($0.2<z<0.6$) are limited to systems with $kT \sim 1$ keV or above, and therefore probe a transitional regime of galaxy associations, which we refer to as massive groups or poor clusters, between what are typically referred to as clusters and lower mass groups of galaxies.

Our group has undertaken a program to study in depth X-ray luminous intermediate-redshift groups/poor clusters selected from the ROSAT Deep Cluster Survey (Rosati et al. 1998).  Spectroscopic confirmation and HST imaging of a sample of nine groups was presented in Mulchaey et al. (2006; Paper I).  Jeltema et al. (2006; Paper II) investigated the X-ray properties of six of these groups based on observations taken with XMM-Newton, and Jeltema et al. (2007; Paper III) presented deeper spectroscopic follow-up of seven groups with the Gemini-North and Keck telescopes.  One of the most interesting results of this study was the discovery that unlike X-ray luminous groups at low redshift, most of these systems either lacked a dominant central galaxy or contained multi-component brightest galaxies indicative of a recent merger (Papers I and III).  In other ways these intermediate-redshift groups are similar to low-redshift groups and clusters; they have X-ray luminosities and temperatures consistent with the low-redshift $L_X-T_X$ relation (Paper II), and they contain high fractions of elliptical galaxies and low fractions of star-forming galaxies (Paper III). These observations are consistent with a picture in which BCGs form/grow late with respect to the collapse of the system itself (De Lucia \& Blaizot 2007; Dubinski 1998).  In fact, the groups in our sample appear to span a range of evolutionary states.  Of the six groups with XMM-Newton observations, two lack a brightest galaxy at the X-ray center and have elongated X-ray emission, two have bright central galaxies composed of three or more luminous nuclei but round X-ray emission, and two have bright galaxies at the X-ray center with no galaxies of similar luminosity nearby.

In this paper, we investigate in detail one particularly intriguing system in our sample, RXJ1648.7+6109 at $z=0.376$.  Based on previous observations, this massive group appears to be in an early state of evolution.  Instead of a single, dominant galaxy it contains a string of seven galaxies within 200 kpc of the X-ray center, five of which are brighter than $L_{\ast}$ (Paper III).  RXJ1648's X-ray emission appears to be elongated, but the XMM-Newton observation was almost completely contaminated by background flares leading to large uncertainties in its X-ray temperature and spatial distribution (Paper II).  Here we report on a new $\sim$ 100 ks Chandra observation of RXJ1648, which allows us to determine its average X-ray properties and spatial structure (\S 3.1).  Combining these new X-ray measurements with the other groups in our sample and intermediate-redshift groups/poor clusters from the literature, we derive the $L_X-T_X$, $L_X-\sigma_v$, and $\sigma_v-T_X$ relations for poor clusters and groups at intermediate redshifts (\S 3.3).  In addition, we investigate in greater detail the galaxy properties in RXJ1648, including the magnitude, star formation, and velocity distributions of the member galaxies (\S 3.2) and the future formation of a BCG through the mergers of the bright galaxies in the central string (\S 4.1).  Throughout the paper, we assume a cosmology of $H_0=70h_{70}$ km s$^{-1}$ Mpc$^{-1}$, $\Omega_m=0.27$, and $\Lambda = 0.73$.

\section{ DATA REDUCTION }

\subsection{ X-ray Data }

RXJ1648.7+6109 was observed twice with ACIS-S in August 2007 for 49.7 ks (ObsID 7903) and 47.6 ks (ObsID 8472).  The data were prepared using CIAO 3.4.1 and CALDB version 3.4.1 following the standard data processing.  We chose to reprocess the data starting from the level 1 file, including re-detecting hot pixels and afterglow events using the latest tools, applying the newest gain file, and destreaking the S4 chip.  We also applied the CTI and time-dependent gain corrections.  We kept only events with \textit{ASCA} grades of 0, 2, 3, 4, and 6 and a status of zero to filter out particle background, bad pixel, cosmic ray afterglow and other ``bad'' events\footnote{\textit{Chandra} Proposers' Observatory Guide http://cxc.harvard.edu/proposer/POG/, section ``ACIS''}.  Since the data were taken in VFAINT (VF) mode, the additional background cleaning for VF mode data was also applied.  We removed background flares from the data following the prescriptions of Markevitch et al. (2003).  The filtering excluded time periods when the count rate, excluding point sources and group emission, was not within 20\% of the quiescent rate.  We detected flares in the 2.5-6 keV band on the back-illuminated CCD, S1 in order to have a significant area free of group emission.  No significant flare periods were found for either observation, and their good exposure times were 49.5 ksec and 47.5 ksec, respectively.

Each observation was reduced separately.  We checked the relative alignment of the two observations using several bright point sources in the field, and found them to be consistent within one pixel (0.5").  Images and exposure maps for each observation were created in the 0.5-2 keV band.  The exposure maps were weighted by an absorbed mekal spectral model with $kT = 1$ keV and galactic absorption.  The individual observation images were then exposure corrected, weighted by their exposure times, and merged.  We extracted spectra for the group, after the removal of point sources, from each observation within the radius where the group flux was above background ($r=75"$) and background spectra from a large annular region surrounding the source region ($r=125"$).  We used the CIAO script specextract to create source and background spectra, RMF and ARF files for both observations.  The source spectra were then grouped to have at least 25 counts per bin.  The earlier XMM observations of this group were almost entirely contaminated by background flaring, and even with a relaxed (i.e.~allowing more background) flare filtering only $\sim 150$ net source counts remained in the spectral extraction region (0.5-8 keV, $S/N \sim 6$).  In contrast, in the new Chandra observations we detect over 650 net source counts (0.5-7 keV, $S/N \sim 11$).

\subsection{ Optical Data }

The optical imaging and spectroscopy as well as the galaxy catalog for this group were presented in Paper III.  Briefly, we observed RXJ1648 with GMOS (Hook et al. 2004) on Gemini-North during May 11-13, 2005 using 4 masks with 12-20 slits per mask and three 30 minute exposures per mask.  GMOS was operated with the B600 G5303 grating with a central wavelength of 5500 \AA\ leading to a spectral coverage of $\sim 3500-7000$ \AA.  Spectra were binned to a pixel scale of 1.8 \AA\ per pixel to improve the signal-to-noise.  The chosen 1'' slit width gives a spectral resolution of approximately 5.5 \AA.  The Gemini spectra were reduced using the Gemini IRAF package.  The reduction included bias subtraction, flat-field correction, cosmic ray rejection, sky subtraction, wavelength calibration, and extraction of one-dimensional spectra.  The 1D spectra of multiple science exposures were combined after the rejection of outlying pixels using the averaged sigma clipping algorithm.

We determine galaxy redshifts using the cross-correlation routine XCSAO (Kurtz \& Mink 1998) using the galaxy and quasar templates from the Sloan Digital Sky Survey (SDSS) Data Release 4.  All redshifts were confirmed by eye by TEJ.  Galaxy morphologies were classified utilizing our HST WFPC2 imaging (F702W filter) where possible or utilizing the Gemini pre-imaging in the r\_G0303 filter.

In this work, we also remeasured the equivalent widths of spectral lines from those listed in Paper III using a slightly modified technique.  Here we consider two lines: [OII] emission as an indication of ongoing star formation, and $H_{\delta}$ absorption as an indication of the Balmer absorption.  We examined the SDSS stellar composite spectra to select continuum bandpasses free of surrounding features; these are listed in Table 1 for the two lines considered.  For $H_{\delta}$ in particular, the continuum features surrounding the line are sensitive to age and metallicity (Dressler et al. 2004), and to avoid features immediately blueward of the line we choose a blue bandpass offset from the line compared to what has been typically used (Balogh et al. 1999).  Using the program viewspectra in the COSMOS package developed by A. Oemler (http://www.ociw.edu/Code/cosmos), the continuum was fit to a straight line between the red and blue bandpasses and the line was fit to a Gaussian.  All fits were checked by eye, and the redshift was adjusted if necessary to correctly select the line center.  In Paper III, for comparison, we derived [OII] equivalent widths through two methods: interactively selecting the continuum as in Dressler et al. (2004) and using a classical bandpass technique, which sums the total flux in the line bandpass rather than performing a fit, with the bandpasses of Balogh et al. (1999).  For strong emission lines like [OII], the difference in technique has little effect; here EW([OII]) increases by on average 1 \AA\ from those used in Paper III because we have slightly improved the line centering.  However, for absorption features like $H_{\delta}$, we found the current method to be more robust to differences in continuum features while also being reproducible.

\section{ RESULTS }

\subsection{ X-ray Properties }

The shape of the X-ray emission of RXJ1648.7+6109 can be seen in Figure 1, which shows the X-ray contours, derived from the adaptively smoothed Chandra image (csmooth), overlaid on the HST WFPC2 image of this group.  The X-ray emission appears to be significantly elliptical and is elongated along the direction of the central string of bright galaxies noted in previous papers (Paper I; Paper III).  A similar X-ray shape was seen in the shallower XMM observation (Paper II), but the contaminated background in this observation meant that we were unable to constrain the surface brightness profile or shape.  The X-ray emission extends over the entire region subtended by this galaxy string and is not clearly peaked on any one of these bright galaxies.  The X-ray peak lies closest in projection to an Sa group member which, however, has the largest velocity offset from the center of the group velocity distribution (910 km/s) and is not one of the brightest galaxies in this region.  The other galaxies in the string all lie close to the center of the velocity distribution, implying that they are truly located in the group core.

\begin{figure}
\epsscale{0.5}
\plotone{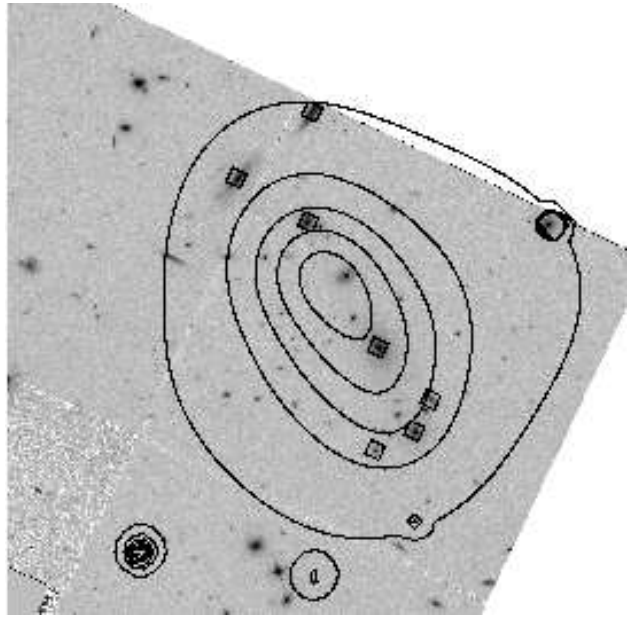}
\caption{ Contours of the smoothed 0.5-2 keV Chandra image overlaid on the \textit{HST} WFPC2 F702W image.  The total exposure of the WFPC2 image is 7800 secs. The seven galaxies lying near the center of the group (both in projected distance and velocity offset) are marked with diamonds. }
\end{figure}

We fit the exposure corrected, X-ray surface brightness distribution to a 2D elliptical $\beta$-model using $Sherpa$.  The best-fit parameters are listed in Table 2.  Similar to our other groups/poor clusters in this redshift range, we find that RXJ1648.7+6109 has a fairly steep surface brightness distribution ($\beta = 1.45^{+1.28}_{-0.40}$) compared to what has been found for low-redshift groups (Paper II; Osmond \& Ponman 2004).  It also has a significant ellipticity of $0.34 \pm 0.07$, similar to two other groups in our intermediate-redshift sample, RXJ0329.0+0256 and RXJ1334.0+3750.  The significant X-ray ellipticity implies that RXJ1648 is dynamically young/unrelaxed; it may still be in the process of collapsing or perhaps has undergone a recent merger.  Overall, half of the six intermediate-redshift groups that we have observed with XMM and Chandra have a major axis at least 50\% larger than the minor axis.  In comparison, only 8 out of 51 (16\%) of the groups with measured ellipticities in the low-redshift sample of Mulchaey et al. (2003) have axis ratios as large, although the groups in the Mulchaey et al. (2003) sample are on average lower mass systems.

To determine the average group temperature, we jointly fit the spectra from the two observations using XSPEC 11.3.2 to an absorbed mekal model.  The spectrum was fit in the 0.5-7 keV band, ignoring the Si K line region.  Combined the two observations contain roughly 650 net group counts in the spectral extraction region.  The group metallicity cannot be well constrained in the current observations, and we fixed this parameter at 0.3 solar (Anders \& Grevessa 1989).  The best-fit spectral model is shown in Figure 2; we find $kT = 2.1^{+0.8}_{-0.6}$ keV and $n_H = 1.5^{+1.1}_{-0.8} \times 10^{21}$ cm$^{-2}$ (compared to a galactic value of $n_H = 2.1 \times 10^{20}$ cm$^{-2}$; Dickey \& Lockman 1990) with $\chi^2 = 95.6/108$ DOF.  This best-fit temperature is higher than the best-fit temperature we previously derived from the XMM data ($kT = 0.97^{+1.59}_{-0.54}$ keV), but consistent with it within the errors (Paper II).  The bolometric (0.01-100 keV) group luminosity within $r_{500}$, the radius at which the group density is 500 times the critical density, is $3.2^{+1.7}_{-0.8} \times 10^{43}$ ergs s$^{-1}$.  We used the best-fit spherical $\beta$-model fit to the surface brightness to extrapolate the luminosity between the spectral extraction radius (385 kpc) and $r_{500}$ (831 kpc), but this correction was less than 4\%.  The average group properties are summarized in Table 2.  As shown in Section 3.3, the luminosity and temperature of RXJ1648 are very consistent with the intermediate-redshift group/poor cluster $L_X-T_X$ relation.  RXJ1648's temperature of $\sim 2$ keV places it in the massive group to poor cluster mass regime, similar to the rest of our RDCS sample.  Unfortunately, X-ray observations of galaxy systems at these redshifts currently do not extend to lower mass groups (see \S 3.3).

\begin{figure}
\epsscale{0.5}
\plotone{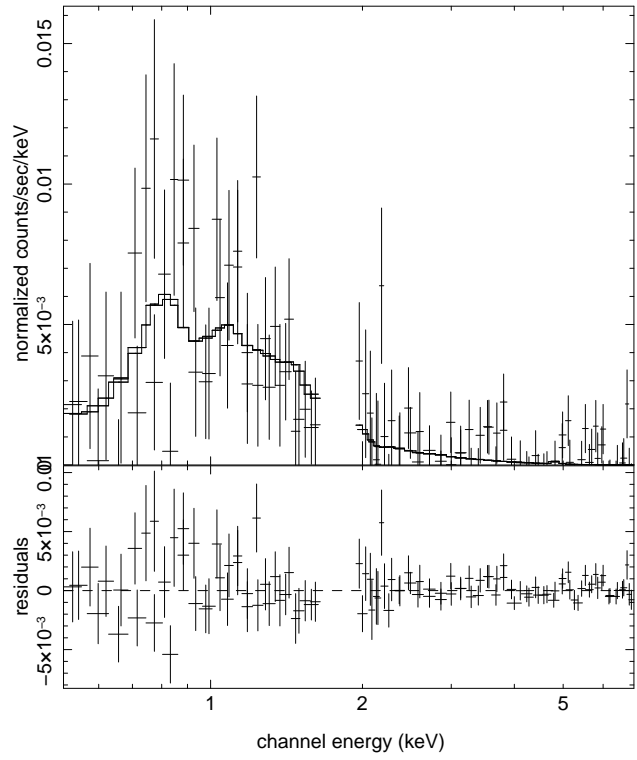}
\caption{ Group spectra from the two Chandra observations and the best-fit absorbed mekal model. }
\end{figure}

\subsection{ Galaxy Properties }

The lack of a central, dominant galaxy in such an X-ray luminous group is very rare at low redshift (Paper I; Osmond \& Ponman 2004), and RXJ1648 appears to be in an early stage of formation before mergers of the member galaxies have formed a central BCG.  As discussed in Paper III, the optical spectroscopy reveals that the bright central string of galaxies in RXJ1648 has a lower velocity dispersion than the group as a whole.  For galaxies with $r<200$ kpc, the velocity dispersion is $194^{+66}_{-23}$ km s$^{-1}$, with one significant outlier excluded, versus $417^{+118}_{-86}$ km s$^{-1}$ for all 22 known group members.  The velocity dispersion was determined using the ROSTAT package (Beers, Flynn, and Gebhardt 1990) using the biweight estimator with errors from bootstrapping.  As discussed in Paper III, we iteratively clip galaxies with velocity offsets from the mean more than 3 times the velocity dispersion.  The velocity distribution for RXJ1648, shown in Figure 3, has a large kurtosis of 3.04, higher than a Gaussian at $>99$\% confidence, and a large Scaled Tail Index of 1.51 (Bird \& Beers 1993), indicating a velocity distribution much broader than a Gaussian.  For example, both a Gaussian with $\sigma = 417$ km s$^{-1}$ and a Gaussian with $\sigma = 194$ km s$^{-1}$ are compared to the actual velocity distribution in Figure 3.  Here $\sigma_v \sim 400$ km s$^{-1}$ over predicts the power in the wings of the distribution while $\sigma_v \sim 200$ km s$^{-1}$ matches well the central velocity distribution but misses the tails at higher velocity offsets.  On the other hand, the velocity distribution is fairly symmetric and does not have a significant skewness or asymmetry index (Bird \& Beers 1993).  Overall, RXJ1648 appears to be composed of a smaller, $200$ km s$^{-1}$ system sitting within a more extended, $400$ km s$^{-1}$ system, suggestive of a virialized core surrounded by an infalling halo, but the statistics are currently limited by the number of known members.  A very similar velocity distribution exhibiting a peaked core component within a higher dispersion halo was found by Bower et al. (1997) for the composite velocity distribution of $z \sim 0.4-0.5$ optically-selected clusters.

\begin{figure}
\epsscale{0.8}
\plotone{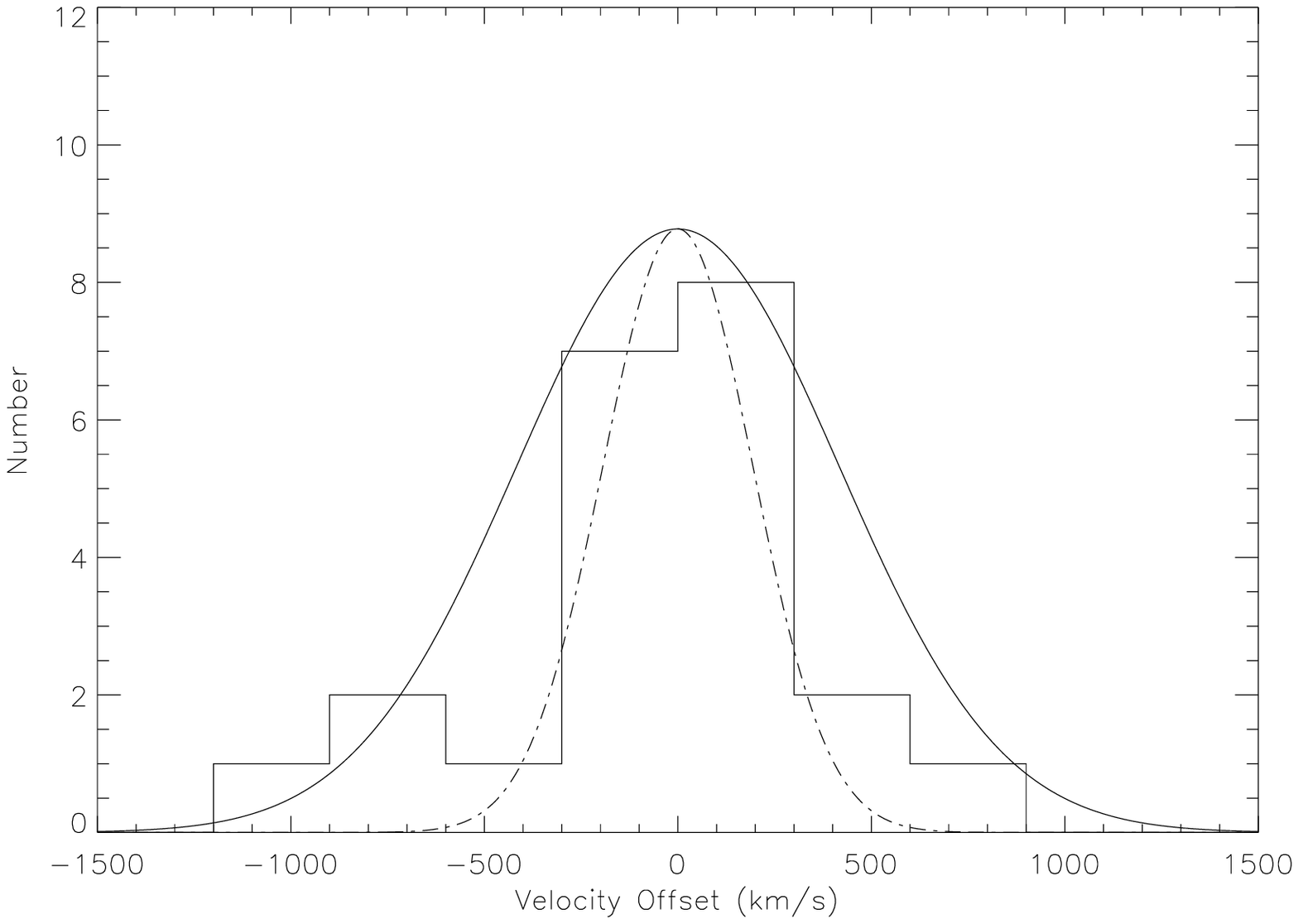}
\caption{ Velocity distribution of RXJ1648 member galaxies relative to the mean velocity of the system.  For comparison, a Gaussian with $\sigma = 417$ km s$^{-1}$ (solid line ) and a Gaussian with $\sigma = 194$ km s$^{-1}$ (dot-dashed line)  are overplotted. }
\end{figure}

We also find significant differences in the properties of the group galaxies with radius.   For example, a rank-sum test shows that galaxies with $r \leq 200$ kpc tend to be brighter ($\langle M_R \rangle = -21.33 \pm 0.06$ for $r > 200$ kpc vs. $\langle M_R \rangle = -21.86 \pm 0.11$ for $r \leq 200$ kpc).  In addition, galaxies with significant [OII] emission (EW([OII]) $> 5$ \AA), indicating ongoing star formation, tend to lie at larger radii (Figure 4b; $\langle r \rangle = 588 \pm 52$ kpc for EW([OII]) $> 5$ \AA\ vs. $\langle r \rangle = 296 \pm 14$ kpc for EW([OII]) $\leq 5$ \AA), and EW[OII] is on average higher for galaxies with large velocity offsets (Figure 4a; $\langle$EW([OII])$\rangle = 7.3 \pm 1.0$ for $v_{off} > 200$ km s$^{-1}$ vs. $\langle$EW([OII])$\rangle = 3.5 \pm 0.9$ for $v_{off} \leq 200$ km s$^{-1}$) from the central group velocity.  This increase in [OII] emission with radius and velocity offset again indicates the presence of an evolved core within a more recently infalling halo.  Similarly, EW($H_{\delta}$), indicating recent star formation ($<1.5$ Gyr), increases with radius (Figure 4c; $\langle$EW($H_{\delta}$)$\rangle = 2.76 \pm 0.11$ for $r > 400$ kpc vs. $\langle$EW($H_{\delta}$)$\rangle = 1.39 \pm 0.09$ for $r \leq 400$ kpc).  Moderate $H_{\delta}$ absorption correlates with significant [OII] emission.  We find one likely post-starburst (k+a) galaxy in RXJ1648, an Sb galaxy in the central string with weak [OII] emission (EW([OII]) $=3.0 \pm 1.1$ \AA), strong Balmer absorption (EW($H_{\delta}$) $=4.6 \pm 0.7$ \AA), and very weak $H_{\beta}$ emission within a strong $H_{\beta}$ absorption feature.  Although, it is only one galaxy, this corresponds to 5\% of the group members for which we have Gemini-GMOS spectra.  One other group galaxy has moderate $H_{\delta}$ absorption (EW($H_{\delta}$) $=3.3 \pm 1.1$ \AA) with no detectable [OII] emission, but the other Balmer absorption features are not strong.  

The average properties of RXJ1648's galaxies are quite similar to the other groups in our sample.  For example, we find that 29\% (6/21) of the group galaxies have significant [OII] emission (EW([OII]) $> 5$ \AA), similar to other intermediate-redshift groups/poor clusters (Paper III); all of these high [OII] galaxies also have detectable $H_{\beta}$ or $H_{\delta}$ emission.  Of the galaxies with significant [OII] emission, we find no evidence for significantly higher or lower [OII] equivalent widths in RXJ1648 versus the other groups in our sample.  Similarly, RXJ1648 has an early-type fraction (E and S0 galaxies) of 71\% very similar to the average of $70^{+7}_{-5}$\% for the total intermediate-redshift group sample (Paper III).  Of the seven galaxies in the central string with low velocity offsets, five are early-type galaxies, one of which is an E0/E1 pair noted in Paper III, with no detectable [OII] emission and EW($H_{\delta}$) $< 1$ \AA, consistent with evolved stellar populations.  The other two central galaxies appear to have slightly younger stellar populations; one of these is the post-starburst Sb galaxy mentioned above, and the other is a sub-$L_{\ast}$ Sa galaxy with EW([OII]) $= 5.0 \pm 0.4$ \AA\ and EW($H_{\delta}$) $=2.0 \pm 1.4$ \AA.  With the exception of these latter two galaxies, the spectral properties of the galaxies in the center of RXJ1648 are very similar to the brightest galaxies of the other intermediate-redshift groups/poor clusters in our sample.  None of these BCGs, including those with multiple nuclei, have detected [OII] emission, and they all have EW($H_{\delta}$) $<2.0$ \AA.

\begin{figure}
\epsscale{0.55}
\plotone{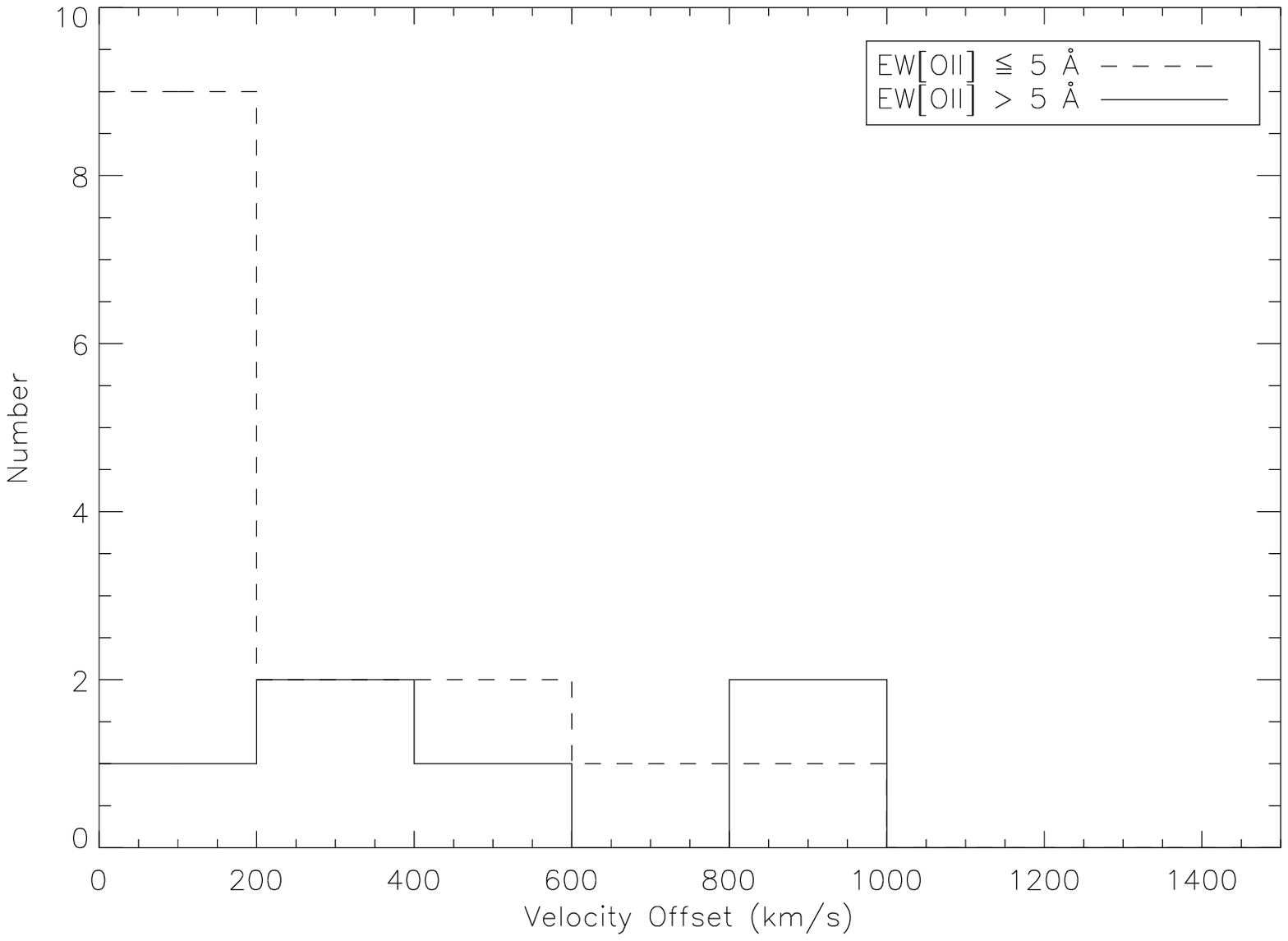}
\plotone{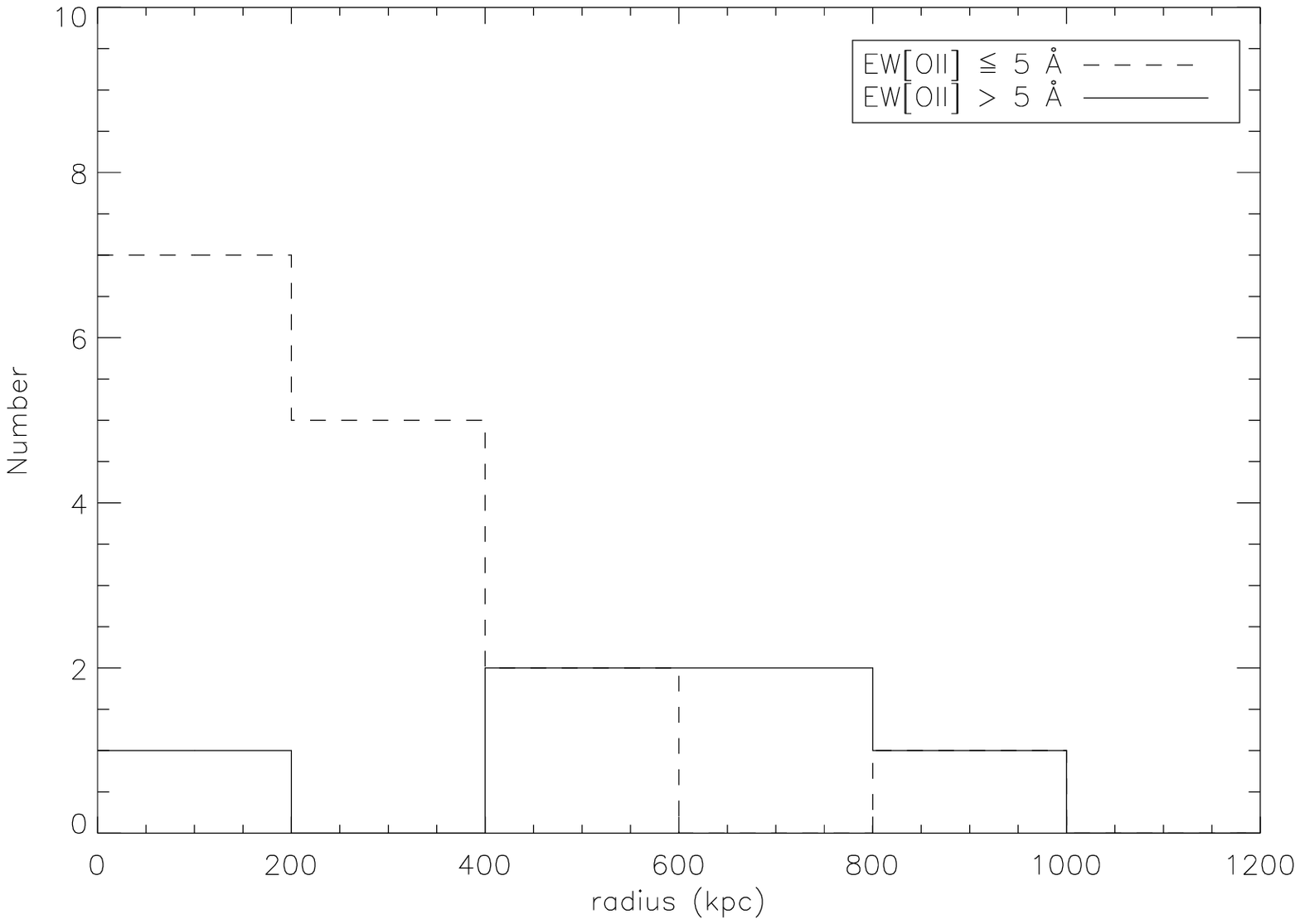}
\plotone{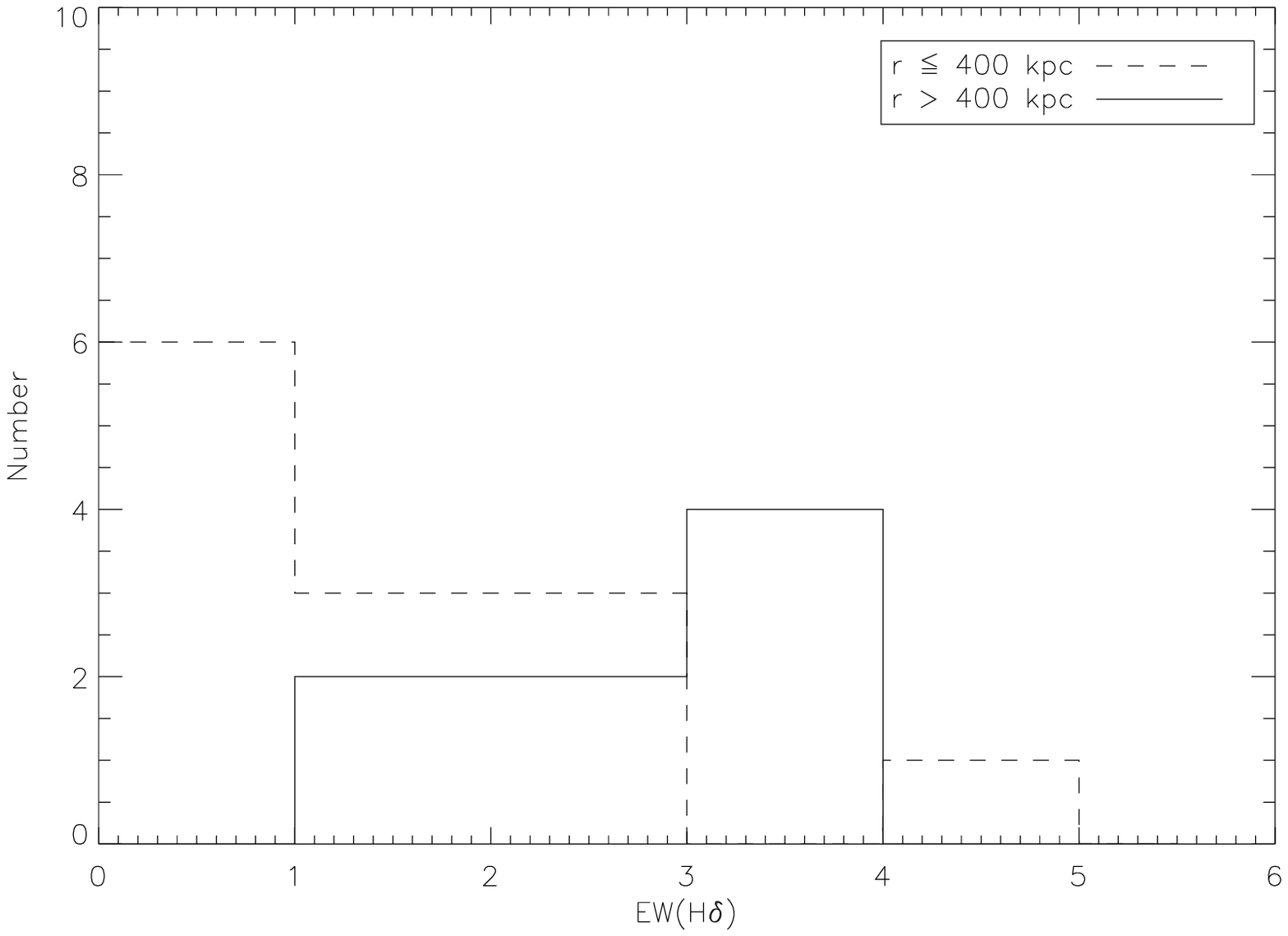}
\caption{ Examples of the distribution of galaxy properties among RXJ1648 member galaxies. Top: Distribution of velocity offsets from the mean group velocity for galaxies with strong (EW([OII]) $> 5$ \AA, solid line) and weak (EW([OII]) $\leq 5$ \AA, dashed line) [OII] emission. Middle: Distribution of radii for galaxies with strong (EW([OII]) $> 5$ \AA, solid line) and weak (EW([OII]) $\leq 5$ \AA, dashed line) [OII] emission. Bottom: Distribution of EW($H_{\delta}$) for galaxies with large ($r > 400$ kpc, solid line) and small ($r \leq 400$ kpc, dashed line) projected radii. }
\end{figure}

Overall, the X-ray and optical data for RXJ1648 support a picture of its formation in which the collapse of the system, indicated by its X-ray luminous, hot IGM, and the transformation of the bulk of the galaxies to early-types with little star formation have occurred before the formation of a dominant BCG.  Similarly, other X-ray luminous groups/poor clusters at intermediate redshifts lack a central galaxy or have multi-component BCGs (Papers I and III).  The generally evolved nature of the galaxies in the central string of RXJ1648 implies that if these galaxies merge to form a BCG these mergers are expected to be ``dry'' (i.e.~not produce significant star formation), as is observed for the two groups in our sample with multiple nuclei BCGs which show no significant star formation.  While the typical galaxy in RXJ1648, particularly in the center of the group, has an early-type morphology and little star formation, both the velocity distribution in this group and the increase in galaxies with significant [OII] emission and $H_{\delta}$ absorption with radius/velocity offset indicate the presence of a younger infalling galaxy population surrounding a more evolved core.

\subsection{ Scaling Relations }

At low-redshift, non-gravitational heating appears to be proportionally more important in the group regime, leading to a steepening of the $L_X-T$ relation for groups versus clusters and excess group entropy above self-similar expectations (e.g.~Helsdon \& Ponman 2000; Ponman et al.~2003).  In particular, the evolution in group scaling relations can distinguish models of non-gravitational heating, for example excess entropy that is constant with redshift (i.e.~preheating) versus being tied to halo cooling time (Balogh et al.~2006).

The scaling relations of groups at intermediate-redshifts are fairly unknown.  Collecting data from the literature, Gastaldello et al. (2007) recently derived some of the first constraints on the $L_X-T_X$, $L_X-\sigma_v$, and $\sigma_v-T_X$ relations for poor clusters and groups at intermediate redshifts.  Here we re-derive these relations with a more restrictive cut on temperature ($kT \leq 3$ keV) and adding the Chandra measurements for RXJ1648, a seventh RDCS group observed with XMM (Mulchaey et al.~2008 in prep.), and additional XMM-LSS groups that were recently published (Pacaud et al. 2007).  Limiting the sample to groups with $0.2<z<0.6$, our sample includes 7 RDCS groups (Paper II; Mulchaey et al.~2008 in prep.), 16 XMM-LSS groups (Pacaud et al. 2007; Willis et al. 2005), and Zw 1305.4+2941 from Gastaldello et al. (2007).  Unfortunately, only 5 of the XMM-LSS groups have published velocity dispersions.  In addition, the current samples at these redshifts are limited to $\sim 1$ keV or hotter systems.  The scaling relations are fit as $log(y) = A + B log(x)$ using the bisector modification to the BCES method of Akritas \& Bershady (1996), which allows for intrinsic scatter and non-uniform measurement errors in both variables.  Errors on the best-fit parameters were determined through 10,000 bootstrap trials.  An $E_z^{-1}$ correction is applied to the X-ray luminosities where $E_z = (\Omega_m(1+z)^3 + 1 - \Omega_{\Lambda})^{1/2}$.  The results are listed in Table 3.

In Figure 5, we compare the $L_X-T_X$ relation for intermediate-redshift groups/poor clusters to the low-redshift group relation from the GEMS sample ($0.2 \leq kT \leq 1.5$; Osmond \& Ponman 2004), as well as to low-redshift clusters ($3.5 \leq kT \leq 9.7$; Markevitch 1998).  RXJ1648 is highlighted in red.  The intermediate-redshift group relation is very similar to the low-redshift cluster relation; it is flatter than the low-redshift group relation, but consistent with it within the errors.  Despite its young dynamical state, RXJ1648 has a luminosity and temperature very consistent with the other groups.

\begin{figure}
\epsscale{0.8}
\plotone{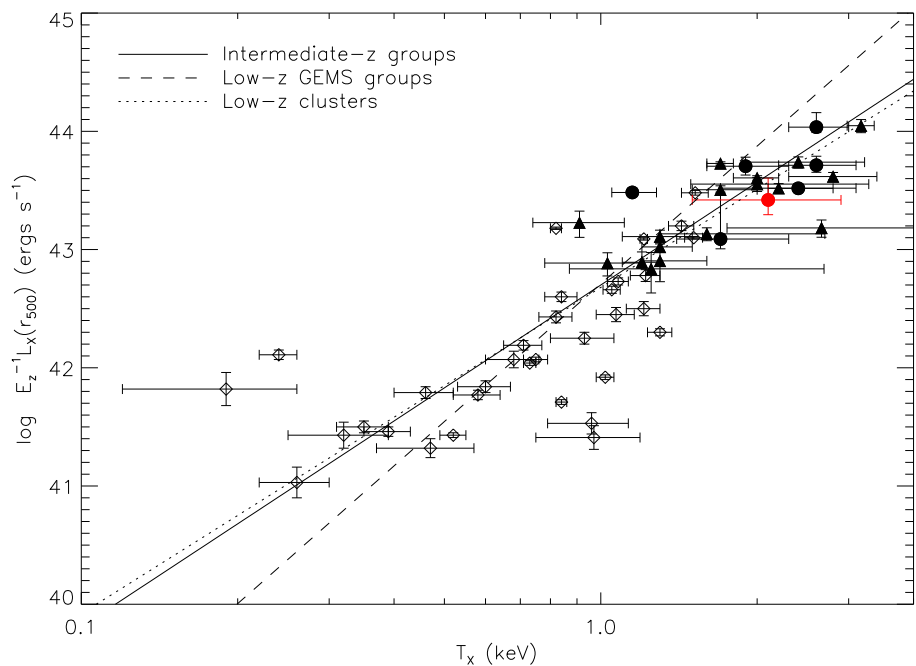}
\caption{ $L_X-T_X$ relation for intermediate-redshift groups/poor clusters (solid line) compared to low-redshift groups (dashed line; Helsdon \& Ponman, in preparation) and low-redshift clusters (dotted line; Markevitch 1998). The RDCS groups are plotted with circles (RXJ1648 in red), and the other intermediate-redshifts groups are plotted with triangles (Pacaud et al. 2007; Willis et al. 2005; Gastaldello et al. 2007).  Also shown is the low-redshift GEMS groups sample (open diamonds; Osmond \& Ponman 2004). }
\end{figure}

The $L_X-\sigma_v$ and $\sigma_v-T_X$ relations are shown in Figure 6.  Here intermediate-redshift groups/poor clusters appear significantly different than low-redshift systems; their velocity dispersions are low compared to their X-ray properties.  Even considering the errors, the intermediate-redshift group $L_X-\sigma_v$ and $\sigma_v-T_X$ relations are significantly different than low-redshift groups and clusters, implying a lower energy in galaxy motions compared to the group gas at higher redshifts.  Figure 6c plots $\beta_{spec} = \mu m_p \sigma_v^2/kT_X$, the ratio of the specific energy in the motion of the galaxies to that in the X-ray emitting gas, for intermediate-redshift groups and poor clusters versus X-ray luminosity.  Here most of the groups have $\beta_{spec} < 1$, some with high significance.  This deviation from $\beta_{spec} = 1$ is not seen in groups at low-redshift at these luminosities (Mulchaey \& Zabludoff 1998; Osmond \& Ponman 2004), although some very low velocity dispersion, low-redshift systems may have $\beta_{spec} < 1$ (Osmond \& Ponman 2004).  We do not observe any trend in $\beta_{spec}$ with $L_X$ for intermediate-redshift groups.

\begin{figure}
\epsscale{0.55}
\plotone{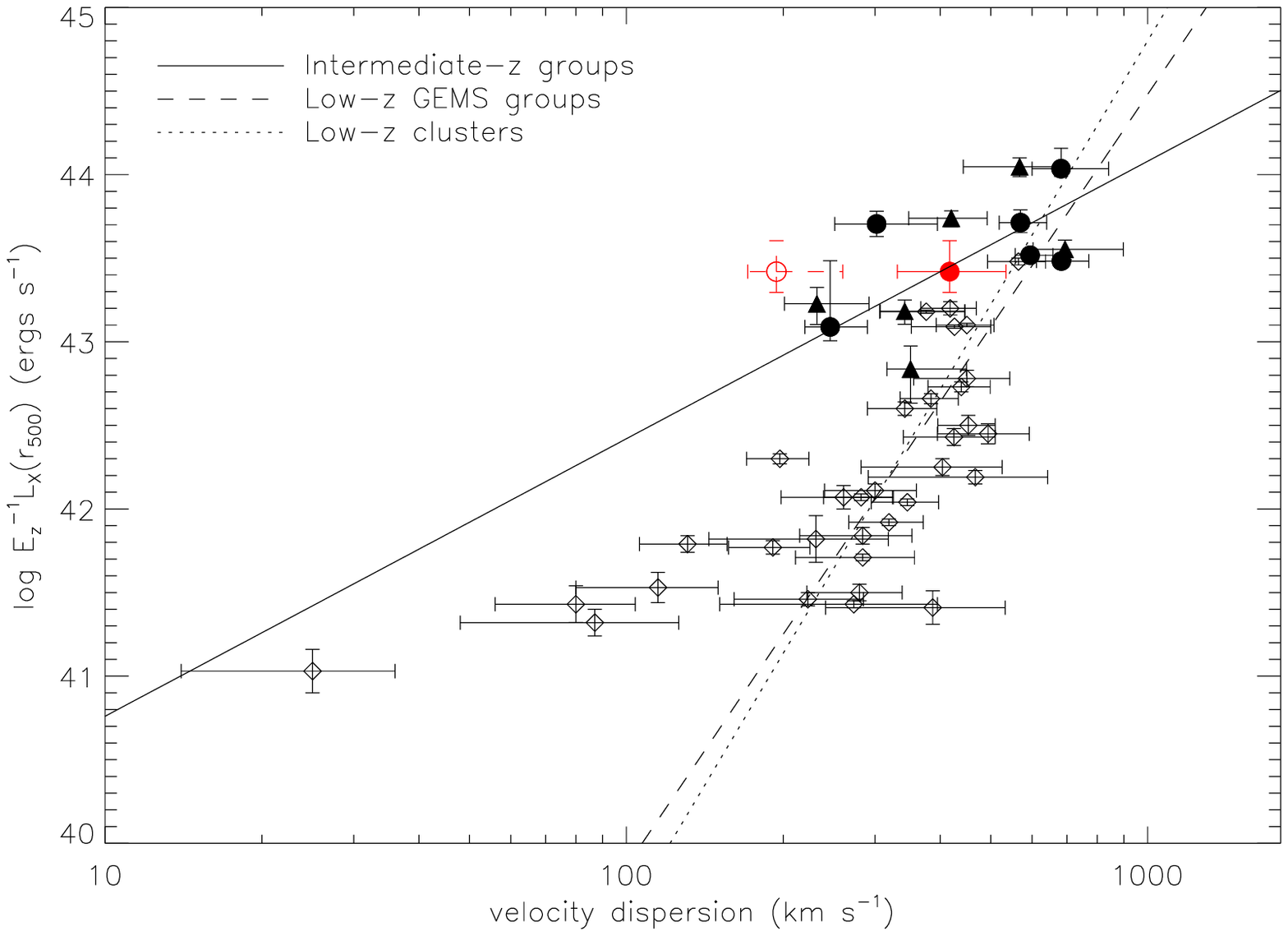}
\plotone{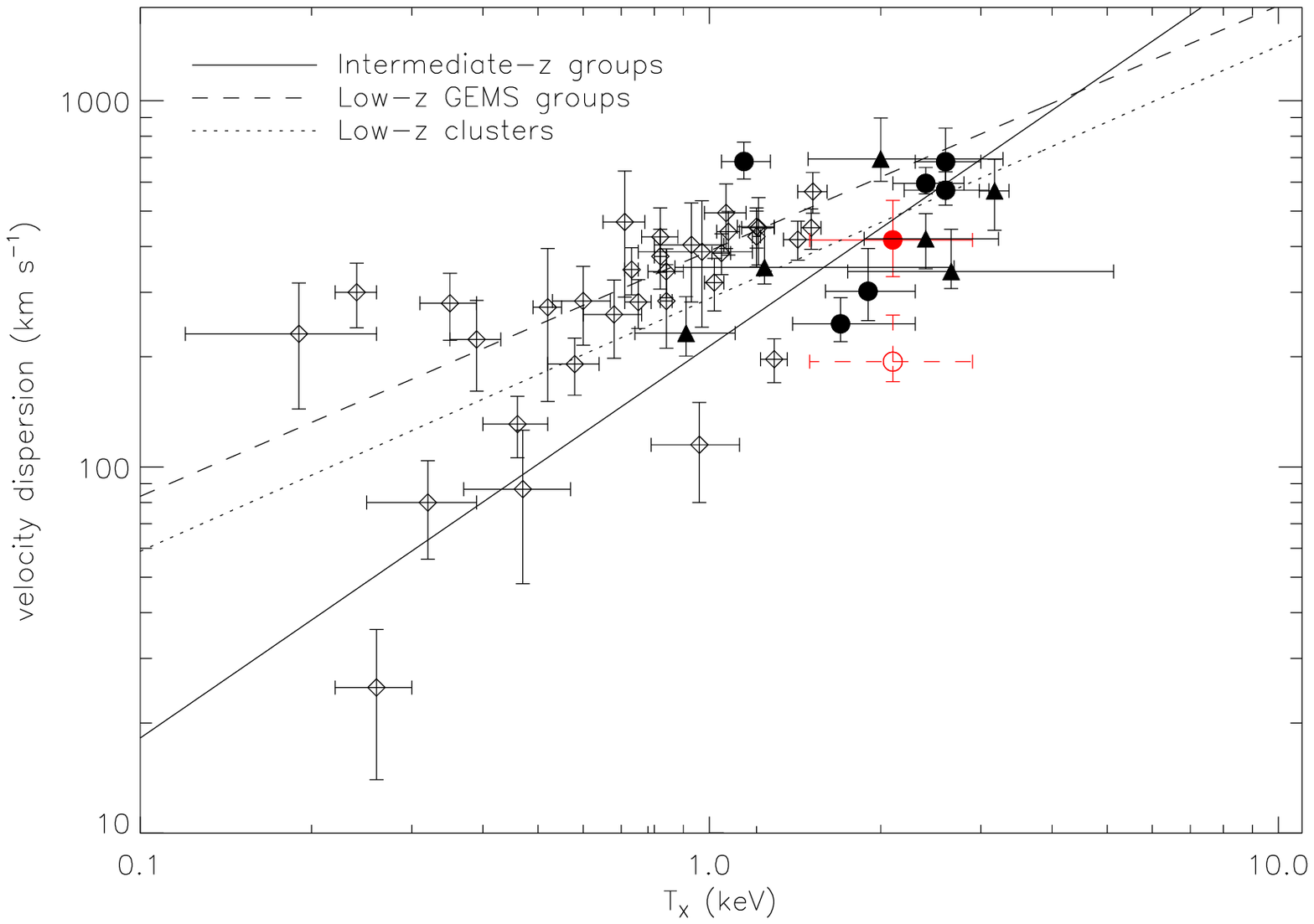}
\plotone{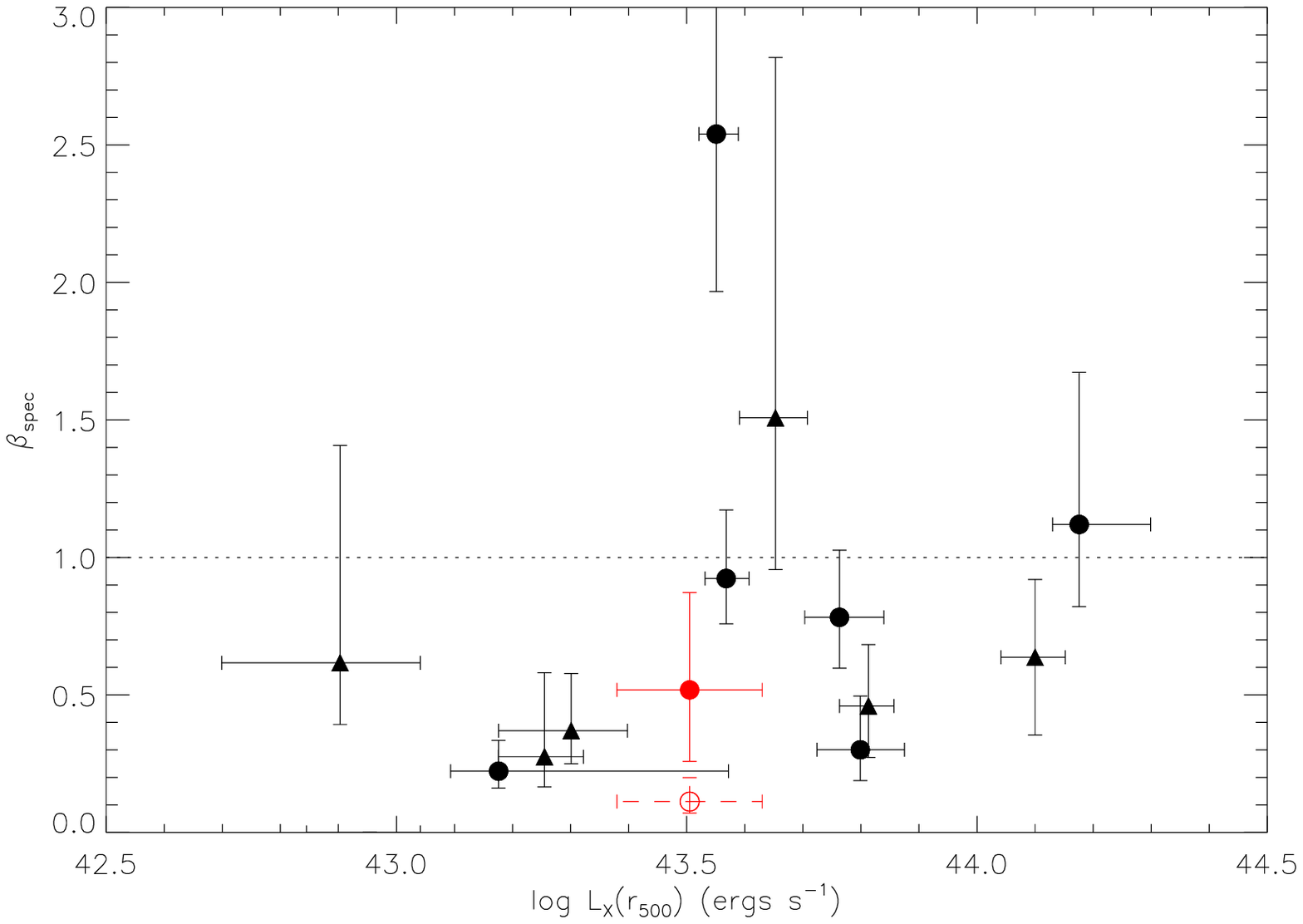}
\caption{ Top: $L_X-\sigma_v$ relation.  Middle: $\sigma_v-T_X$ relation.  Bottom: $\beta_{spec}$ versus $L_X$. Symbols are the same as in Figure 5.  For illustration, we plot the central velocity dispersion of RXJ1648 with an open circle and dashed error bars. }
\end{figure}

Why might the velocity dispersions be low compared to low-redshift groups?  Velocity dispersions based on a few of the brightest members can be biased (Zabludoff \& Mulchaey 1998).  In fact, early estimates of the velocity dispersion of RXJ1648 based on eight members, primarily the central bright galaxy string, were very low (Paper I).  However, adding members can also have the effect of lowering the measured velocity dispersions of groups and clusters, because with better sampling more outliers or infalling galaxies tend to be excluded (Gal et al. 2008).  With the current spectroscopy, we note that all of our RDCS groups are fairly well sampled (10-30 members; Paper III), and the intermediate-redshift groups from the literature all have at least eight members. In comparison, many of the low-redshift groups are less well sampled (Osmond \& Ponman 2004).  Another possible bias in the current intermediate-redshift sample is the small luminosity range, which both limits the baseline over which we measure the slope of the scaling relations and may mean that we miss lower luminosity groups in the same velocity dispersion range.  For example, the flux limit of the deepest portion of the RDCS survey is $10^{-14}$ ergs cm$^{-2}$ s$^{-1}$, which corresponds to roughly a bolometric luminosity of $3 \times 10^{42}$ ergs s$^{-1}$ at $z=0.2$ and $3 \times 10^{43}$ ergs s$^{-1}$ at $z=0.6$, and for the most part our XMM data samples the higher-redshift/higher-luminosity objects (Paper I).  In addition, the intermediate-redshift groups are all X-ray selected compared to the heterogeneous selection of the GEMS sample.  Observations of lower luminosity, intermediate-redshift groups would help to resolve these questions, but requires long X-ray observations and significant optical spectroscopy.

For the specific case of RXJ1648 (and possibly 1-2 of the other RDCS groups), the ellipticity of the X-ray emission and lack of a BCG indicate that this system is dynamically young.  Therefore, the gas and galaxies may not be virialized and tracing the same potential, or a past merger could have boosted the X-ray luminosity and temperature (e.g. Rowley et al. 2004; Ricker \& Sarazin 2001).  For illustration, we also plot in Figure 6 the location of RXJ1648 on the scaling relations if we used the velocity dispersion of the central galaxies ($r<200$ kpc) instead of the global velocity dispersion.  With this lower velocity dispersion, RXJ1648 would fall significantly off the scaling relations for the other intermediate-redshift groups and have the lowest $\beta_{spec}$ in the sample.  While the shape and position of the X-ray emission traces the bright, central string of galaxies, the X-ray luminosity and temperature are a better reflection of the dynamics of the global $\sim 400$ km s$^{-1}$ system.

\section{ Discussion }

The X-ray luminosity and temperature of RXJ1648 ($L_{X,bol}(r<r_{500}) = 3.2^{+1.7}_{-0.8} \times 10^{43}$ ergs s$^{-1}$, $kT = 2.1^{+0.8}_{-0.6}$ keV) indicate that that it is a massive, collapsed group.  In addition, the average group properties agree well with the intermediate-redshift group/poor cluster scaling relations.  However, the X-ray emission is very elliptical/elongated, and RXJ1648 lacks a dominant BCG at the X-ray center, indicating that this group is still in the process of formation.  Instead, RXJ1648 contains a string of seven galaxies, five of which are brighter than $L_{\ast}$, within a projected radius of 200 kpc (excluding one galaxy with a velocity offset $>900$ km s$^{-1}$).  The X-ray emission and its elongation trace this central string of galaxies.  These central galaxies have a smaller velocity dispersion, are on average brighter, and have less star formation (lower [OII] and $H_{\delta}$) than other group galaxies.  The total group velocity dispersion of $417^{+118}_{-86}$ km s$^{-1}$ also indicates a massive group and matches well the X-ray luminosity and temperature; however, the velocity distribution indicates a double Gaussian or large tails, with one system having $\sigma \sim 200$ km s$^{-1}$ (dominated by the central string) within a larger system.  

Overall, the observations support the picture that RXJ1648 is a recently formed massive group.  Several of the bright galaxies appear to have sunk towards the center of the system, but have not yet merged to form a dominant, central BCG.  The groups in our X-ray selected, intermediate-redshift sample represent an evolutionary sequence in BCG formation, with RXJ1648 representing one of the youngest systems before a dominant BCG has formed.  Two other RDCS groups, both at $z=0.23$, contain multi-component dominant central galaxies with at least three nuclei, indicating a past merger, and round, relaxed X-ray morphologies (Paper III; Paper I).  These groups represent what RXJ1648 might look like in a few Gyrs.  The formation of a central, dominant BCG through mergers of bright group galaxies appears to occur after group collapse and the appearance of luminous X-ray emission.  Below we discuss more quantitatively the prospects for the formation of a central BCG in RXJ1648.

\subsection{ BCG Formation }

RXJ1648 has a total mass of $M_{200} = 4^{+3}_{-2} \times 10^{14} M_{\odot}$, where the mass is estimated from the X-ray temperature and the surface brightness profile assuming hydrostatic equilibrium.  For a group of this mass, we expect a BCG with an absolute R-band magnitude between $-23.5$ and $-24$ (Brough et al. 2008; Hansen et al. 2008).  The brightest observed galaxy in RXJ1648 is about a magnitude fainter than this prediction with $M_R = -22.8$.  Will the central string of galaxies in RXJ1648 merge, and if so will the resulting galaxy have a luminosity consistent with the total group mass?

First, we consider the time it will take for the galaxies in the string to sink to the group center through dynamical friction.  Using the Faber-Jackson relation (Bernardi et al. 2005), we estimate that the five super-$L_{\ast}$ galaxies within $r<200$ kpc have stellar velocity dispersion between 265 and 350 km s$^{-1}$.  Following the methodology in Zabludoff \& Mulchaey (1998), we find that these galaxies have predicted tidal radii of $\sim 80$ kpc ($r_T \approx r_c \sigma_{gal}/2\sigma_{grp}$ where $r_c$ is the group core radius and $\sigma_{gal}$ and $\sigma_{grp}$ are the galaxy stellar velocity dispersion and the group velocity dispersion, respectively; Merritt 1984) and masses between $6 \times 10^{11} M_{\odot}$ and $14 \times 10^{11} M_{\odot}$ ($M \approx r_T\sigma_{gal}^2/2G$).  If we instead conservatively assume $M/L \approx 5 M_{\odot}/L_{\odot}$, we find galaxy masses about a factor of three lower, but these lead to a similar range of dynamical friction times.  We calculate the time it will take for each galaxy to fall to the center of the group through dynamical friction (eq.[7-18] and eq.[7-13b] in Binny \& Tremaine 1987) approximating the group as a singular isothermal sphere with the total group velocity dispersion (eq.[4-123] in Binny \& Tremaine 1987) and assuming that the galaxies in the string are currently orbiting with a velocity of $\sim 200$ km s$^{-1}$ at their projected distances.  We find that the five super-$L_{\ast}$ galaxies will reach the center in between 1-5 Gyrs (regardless of our definition of galaxy mass).  The group redshift corresponds to 4.1 Gyrs ago, in which time four of the galaxies should merge.  The two sub-$L_{\ast}$ galaxies in the string will not reach the center for $\sim 9$ Gyrs.  These calculations represent a fairly simple estimate of the dynamical friction timescale ignoring, for example, the gravitational affect of the galaxies on each other, but they do show that it is likely that some of these galaxies will merge by the present epoch.

If the five bright string galaxies merged and no stars were lost in the merging process the resulting BCG would have $M_R = -24.23$, a bit brighter then predicted for a group with the mass of RXJ1648.  If only the four brightest galaxies merge, the BCG magnitude would still be $M_R = -24.06$.  Recently, various groups have argued based on comparisons between simulations and observations that a significant fraction of stars (30-80\%) in galaxy mergers within clusters are lost to the intracluster light (ICL) (Conroy, Wechsler, \& Kravtsov 2007; Monaco et al.~2006).  Assuming the observed $L_{BCG}-M_{200}$ relation (Brough et al. 2008; Hansen et al. 2008), we similarly find that 20-50\% of the stars should be lost in the mergers of the five bright string galaxies in order to produce a BCG with the right luminosity for the group mass.  However, if RXJ1648's mass grows with time or fewer galaxies merge, than no loss of stars to the ICL is necessary.

We note that if the bright, central galaxies in RXJ1648 merge, these mergers are expected to be dry.  Of the five super-$L_{\ast}$ galaxies in the central region all are elliptical or S0 galaxies with no detected [OII] emission except for the post-starburst, Sb galaxy noted in \S3.2.  As noted in Paper III, one of these galaxies is in fact a merging system composed of two elliptical galaxies.

\section{ Summary }

1) \textit{RXJ1648.7+6109:}  This massive group was previously identified from XMM and HST imaging as a dynamically young group at $z=0.376$ (Paper II; Paper III), but the XMM observation was highly contaminated by background flares.  Here we re-observed RXJ1648 with Chandra to determine its X-ray temperature and structure.  RXJ1648 is a massive group with a Chandra temperature of $2.1^{+0.8}_{-0.6}$ keV that appears to have recently collapsed.  While this group follows the scaling relations ($L_X-T_X$, $L_X-\sigma_v$ and $\sigma_v-T_X$) of other intermediate-redshift groups/poor clusters, the X-ray emission is significantly elongated and it lacks a central, dominant BCG.  Instead, the elongation of the X-ray emission traces a string of bright galaxies with projected positions near the center of the group.  Our GMOS spectroscopy reveals that these central galaxies have a lower velocity dispersion than the group as a whole, consistent with having sunk toward the group center.  They are also on average brighter and have less [OII] emission than other group galaxies.  We find that the 4-5 brightest galaxies in this string should sink to the center and merge through dynamical friction by the present epoch forming a BCG consistent with or brighter than predicted for a group of RXJ1648's mass.  If these galaxies merge, up to 50\% of the light could be lost to the ICL while still producing a BCG of the right magnitude.  RXJ1648 appears to be in a unique stage of formation: its high X-ray luminosity and temperature indicate a collapsed system, and the bulk of its galaxies have early-type morphologies and little star formation, typical of a dense environment; however, both the velocity distribution in this group and the increase in galaxies with significant [OII] emission and $H_{\delta}$ absorption with radius/velocity offset indicate the presence of a younger infalling galaxy population surrounding a more evolved core, and a central BCG has not yet formed.

2) \textit{Intermediate-redshift groups:}  In total we have deep X-ray observations and spectroscopy for seven $0.2<z<0.6$ groups/poor clusters selected from the RDCS.  We combine this sample with other X-ray selected, intermediate-redshift groups/poor clusters from the literature (Pacaud et al. 2007; Willis et al. 2005; Gastaldello et al. 2007) to fit the $L_X-T_X$, $L_X-\sigma_v$ and $\sigma_v-T_X$ scaling relations, and we compare these to the scaling relations of low-redshift groups and clusters.  The $L_X-T_X$ relation for intermediate-redshift groups/poor clusters is very similar to the low-redshift cluster relation and consistent with the low-redshift group relation, though flatter, within the errors.  In contrast, the $L_X-\sigma_v$ and $\sigma_v-T_X$ relations are significantly different than low-redshift groups and clusters.  The velocity dispersions of intermediate-redshift groups are low compared to their X-ray properties, implying a lower energy in galaxy motions compared to the group gas at higher redshifts.  However, the current sample of intermediate-redshift groups with measured velocity dispersions is small (13 groups) and limited to fairly high X-ray luminosities of $\sim 10^{43}$ ergs s$^{-1}$ or greater.  Pushing down in luminosity at these redshifts would help to both probe for a break in the $L_X-T_X$ relation and to confirm an evolution in the relations between velocity dispersion and X-ray luminosity and temperature.

\acknowledgments
We would like to thank the referee for their insightful comments on our paper.  We would also like to sincerely thank A. Oemler for all of his help, specifically modifying viewspectra for the equivalent width fitting, and S. Profumo for useful discussions.  Support for this work was provided by the National Aeronautics and Space Administration through Chandra Award Number G07-8123X issued by the Chandra X-ray Observatory Center and through NASA grant NNG04GC846.  T.E.J. is also grateful for support from the Alexander F. Morrison Fellowship, administered through the University of California Observatories and the Regents of the University of California. Based on observations obtained at the Gemini Observatory, which is operated by the Association of Universities for Research in Astronomy, Inc., under a cooperative agreement with the NSF on behalf of the Gemini partnership: the National Science Foundation (United States), the Particle Physics and Astronomy Research Council (United Kingdom), the National Research Council (Canada), CONICYT (Chile), the Australian Research Council (Australia), CNPq (Brazil), and CONICET (Argentina).

\begin{deluxetable}{lccc}
\tabletypesize{\small}
\tablecaption{ Equivalent Width Continuum and Line Regions }
\tablewidth{0pt}
\tablecolumns{4}
\tablehead{
\colhead{Line} & \colhead{Line Band} & \colhead{Blue Continuum Band} & \colhead{Red Continuum Band}
}
\startdata
[OII] &3717.0-3737.0 &3680.0-3710.0 &3745.0-3775.0\\
$H_{\delta}$ &4081.0-4121.0 &4000.0-4040.0 &4121.0-4165.0\\
\enddata
\end{deluxetable}

\newpage
\begin{deluxetable}{lc}
\tablecaption{ RXJ1648.7+6109 Properties }
\tablewidth{0pt}
\tablecolumns{2}
\tablehead{
\colhead{}
}
\startdata
$z$ &0.376 \\
$L_{X,bol}(r<r_{500})$ &$3.2^{+1.7}_{-0.8} \times 10^{43}$ ergs s$^{-1}$ \\
$kT$ &$2.1^{+0.8}_{-0.6}$ keV \\
$\sigma_v$ &$417^{+118}_{-86}$ km s$^{-1}$\\
$r_{500}$ &831 kpc \\
$r_c$ & $58"^{+29}_{-13}$  \\
$\beta$ & $1.45^{+1.28}_{-0.40}$ \\
$\epsilon$ & $0.34 \pm 0.07$ \\
\enddata
\end{deluxetable}

\begin{deluxetable}{lcccc}
\tabletypesize{\small}
\tablecaption{ Intermediate-Redshift Scaling Relations }
\tablewidth{0pt}
\tablecolumns{5}
\tablehead{
\colhead{Relation} & \colhead{B} & \colhead{A} & \colhead{scatter} & \colhead{Number of Groups}
}
\startdata
$E_z^{-1}L_X-T_X$ &$2.9 \pm 0.7$ &$42.7 \pm 0.2$ &0.29 &24 \\
$E_z^{-1}L_X-\sigma_v$ &$1.7 \pm 0.4$ &$39.1 \pm 1.0$ &0.28 &13 \\
$\sigma_v-T_X$ &$1.1 \pm 1.0$ &$2.3 \pm 0.3$ &0.18 &13 \\
\enddata
\tablecomments{ Best-fit scaling relations in the form of $log(y) = A + B log(x)$ for intermediate-redshift groups.  The number of groups included in each fit in listed on column 5.  Scatter is calculated as $[ \sum_i (log(y_i) - B - Alog(x_i))^2/N ]^{1/2}$. }
\end{deluxetable}

\end{document}